\documentclass[conference]{IEEEtran}
\usepackage{amsmath}
\usepackage{graphicx}
\usepackage{amssymb}
\usepackage[english]{babel}
\usepackage{multirow}
\usepackage{booktabs}
\usepackage{mathtools}
\usepackage{subcaption}
\usepackage{relsize}
\usepackage[usenames, dvipsnames]{color}
\usepackage{booktabs}
\newcommand{\tabitem}{~~\llap{\textbullet}~~}
\usepackage{enumitem}
\usepackage{lineno}
\usepackage[colorlinks = true,
            linkcolor = blue,
            urlcolor  = blue,
            citecolor = red,
            anchorcolor = blue]{hyperref}

\title{Inter- and Intra-Patient ECG Heartbeat Classification For Arrhythmia Detection: A Sequence to Sequence Deep Learning Approach}

\onecolumn
\author{\IEEEauthorblockN{Sajad Mousavi, Fatemeh Afghah}\\ \vspace{-0.7cm}
\IEEEauthorblockA{\textit{School of Informatics, Computing and Cyber Systems}, 
{Northern Arizona  University}, Flagstaff, USA \\ \{SajadMousavi,Fatemeh.Afghah\}@nau.edu}\\ 
\vspace{-0.5cm}
\IEEEauthorblockN{U. Rajendra Acharya}\\  \vspace{-0.7cm}
\IEEEauthorblockA{\textit{Department of Electronics and Computer Engineering, Ngee Ann Polytechnic, Singapore}
\\
\textit{Department of Biomedical Engineering, School of Science and Technology, Singapore University of Social Sciences, Singapore}\\
\textit{Department of Biomedical Engineering, Faculty of Engineering, University of Malaya, Malaysia} \\
aru@np.edu.sg} 
}

\begin{document}
%
\maketitle

\begin{abstract} 

Electrocardiogram (ECG) signal is a common and powerful tool to study heart function and diagnose several abnormal arrhythmias. While there have been remarkable improvements in cardiac arrhythmia classification methods,  they still cannot offer acceptable performance in detecting different heart conditions, especially when dealing with imbalanced datasets. In this paper, we propose a solution to address this limitation of current classification approaches by developing an automatic heartbeat classification method using deep convolutional neural networks and sequence to sequence models. We evaluated the proposed method on the MIT-BIH arrhythmia database, considering the intra-patient and inter-patient paradigms, and the AAMI EC57 standard. The evaluation results for both paradigms show that our method achieves the best performance in the literature (a positive predictive value of 96.46\% and sensitivity of 100\% for the category S, and a positive predictive value of 98.68\% and sensitivity of 97.40\%  for the category F for the intra-patient scheme; a positive predictive value of 92.57\%  and sensitivity of 88.94\% for the category S, and a positive predictive value of 99.50\% and sensitivity of 99.94\%  for the category V for the inter-patient scheme.). The source code is available at \url{https://github.com/SajadMo/ECG-Heartbeat-Classification-seq2seq-model} \footnote{This material is based upon work supported by the National Science Foundation under Grant Number 1657260. Research
reported in this publication was supported by the National Institute On Minority Health And Health
Disparities of the National Institutes of Health under Award Number U54MD012388.}. 
\end{abstract}


%
\begin{IEEEkeywords}
ECG analysis, heartbeat classification, deep learning, sequence to sequence model, RNNs. 
\end{IEEEkeywords}
\section{Introduction}
\label{sec:intro}
An Electrocardiogram (ECG) is a common non-invasive tool to record heart activities and detect different abnormalities in heart functionality. Classification of the arrhythmic heartbeats in the ECG signal can be a challenging and time-consuming task for a physician, therefore, such heartbeat hand-annotating is often prone to error. This calls for automatic heartbeat classification methods that are able to diagnose arrhythmic heartbeats in real-time with high accuracy.
Several machine learning algorithms such as support vector machines (SVM), multilayer perceptron (MLP), reservoir computing with logistic regression (RC) and decision trees have been utilized for arrhythmia detection \cite{ye2010arrhythmia,escalona2015electrocardiogram,zaeri2018feature,yu2008integration,afghah2015shapley,miran2018real,chen2018predictive,chen2017remote}. These shallow machine learning methods for ECG processing usually follow three main steps, including 1) signal pre-processing, which includes noise removal methods, heartbeat segmentation, etc; 2) feature extraction; and 3) learning/classification. Even though these methods with hand-engineered features and applying noise removal techniques have achieved acceptable performances, deep learning approaches (i.e., automated feature extractions) have shown impressive  results in various domains ranging from computer vision and reinforcement learning to natural language processing \cite{sutskever2014sequence,mousavi2017traffic,mousavi2016learning,mousavi2016deep,mousavi2017applying,shamsoshoara2018distributed,mousavi2014automatic,moghaddam2012learning} as well as more applicable outcomes in biomedical signal processing \cite{kachuee2018ecg,acharya2017deep,mousavi2018ecgnet,mousavi2019sleepeegnet}.         

One of the main limitations of the current heartbeat classification methods including shallow and deep machine learning methods is their poor performance when dealing with imbalanced datasets. In particular, they attain a low positive predictive value and sensitivity for the classes with lower sample size in the dataset. For instance, the majority of existing ECG analysis techniques achieve a low sensitivity in the MIT-BIH arrhythmia database for ventricular escape beat (S) and fusion of ventricular, and normal beat (F) classes. Furthermore, most previously reported works in the literature have been evaluated based on intra-patient paradigm rather than the inter-patient scheme which is an obviously more realistic scenario to prevent training and test the model using the samples from the same patients. Therefore, although some of these methods achieved good accuracies using the intra-patient scheme, their results are not reliable as their evaluation process was biased \cite{de2004automatic}. 

As mentioned above, the conventional arrhythmia classification systems can be generally divided into two categories of inter-patient and intra-patient paradigms in terms of their evaluation mechanism. In intra-patient paradigm, the training and evaluation datasets can include heartbeats from the same patients, while in inter-patient paradigm, a more realistic evaluation mechanism is used where the heartbeat sets for test and training come from different individuals. One of our aims in this paper is to evaluate the proposed method with both the paradigms.

Inspired by the aforementioned issues with the previous works, this paper proposes a novel and effective approach for automatic ECG-based heartbeat classification by leveraging a sequence to sequence deep learning method and an oversampling method named Synthetic Minority Over-sampling Technique (SMOTE) to address the aforementioned challenge with minority classes. The proposed model is evaluated using inter-patient and intra-patient paradigms where it achieves the best results compared to the existing works in the literature.

The rest of this paper is organized as follows. Section \ref{sec:dataset} introduces the database utilized in this study. Section \ref{sec:propsed} describes the proposed method. Section \ref{sec:experimental} presents the experimental setup and shows the achieved results by the proposed method along with a performance comparison to the state-of-the-art algorithms. Finally, Section \ref{sec:conc} concludes the paper. 

\section{Dataset} 
\label{sec:dataset}
In this study, we used the PhysioNet MIT-BIH Arrhythmia database to evaluate the performance of our proposed method \cite{PhysioNetmitdb,moody2001impact}. The MIT-BIH dataset includes the ECG signals for 48 different subjects recorded at the sampling rate of 360Hz. Each record contains two ECG leads; ECG lead II and lead V1. Usually, the lead II is used to detect heartbeats in the literature. Similarly, here in all experiments, we have applied ECG lead II. This database is recommended by the American association of medical instrumentation (AAMI) \cite{ec571998testing}, since it includes the five essential arrhythmia groups as described in Table \ref{tab:cat_beats}.

\begin{table}[h]  
\caption{ Categories of heartbeats existed in the
MIT-BIH database based on AAMI.}
 \centering{
\label{tab:cat_beats}
	\resizebox{0.5\linewidth}{!}{  
\begin{tabular}{c|l}
\toprule
\textbf{Category} &  \textbf{Class}\\
\midrule
 \multirow{4}{*}{\textbf{N}} &  \tabitem Normal beat (N)\\
 &\tabitem Left and right bundle branch block beats (L,R) \\
  &\tabitem Atrial escape beat (e) \\
    &\tabitem Nodal (junctional) escape beat (j) \\
    \\
     \multirow{4}{*}{\textbf{S}} & \tabitem Atrial premature beat (A)\\
 &\tabitem Aberrated atrial premature beat (a) \\
  &\tabitem Nodal (junctional) premature beat (J) \\
    &\tabitem Supraventricular premature beat (S) \\
    
        \\
     \multirow{2}{*}{\textbf{V}} & \tabitem Premature ventricular contraction (V)\\
 &\tabitem Ventricular escape beat (E) \\
\\
\multirow{2}{*}{\textbf{F}} & \tabitem Fusion of ventricular and normal beat (F)\\
\\
\multirow{3}{*}{\textbf{Q}} & \tabitem Paced beat (/)\\
 & \tabitem Fusion of paced and normal beat (f)\\
  & \tabitem Unclassifiable beat (U)\\

 \bottomrule  
\end{tabular} }
}
\end{table}

We considered two main paradigms of inter-patient and intra-patient to evaluate the proposed model. In the intra-patient paradigm, two sets of data samples (beats) are chosen randomly as training and test samples in which there may be the heartbeat samples of the same patient in the training and test sets. While, in the inter-patient paradigm, the training and test set are constructed from different patients, following the protocol proposed by de Chazal et al. \cite{de2004automatic}. In this method, the heartbeats from the MIT-BIH database (44 records based on AAMI) are divided into two sets of records: DS1 = \{101, 101, 106, 108, 109, 112, 114, 115, 116, 118, 119, 122, 124, 201, 203, 205, 207, 208, 209, 215, 220, 223,230\} and DS2 = \{ 100, 103, 105, 111, 113, 117, 121, 123, 200, 202, 210, 212, 213, 214, 219, 221, 222, 228, 231, 232, 233, 234\}. DS1 is used to build the classification model and DS2 is utilized to test the model. Using this division approach, there is no concern about including the heartbeats from the same patient in both training and test sets.

\section{Methodology}
\label{sec:propsed}
In the following sections, we present a detailed description of our proposed novel model to automatically classify each heartbeat of a given ECG signal. 

\subsection{Pre-processing}
The input of this method is a sequence of ECG beats. In order to extract the ECG beats from a given ECG signal, we follow a few simple steps:

\begin{enumerate}[noitemsep,nolistsep]
	\item Normalizing the given ECG signal to the range of between
zero and one. 
	\item Finding the set of $t$ waves regarding the ECG R-peaks of its corresponding annotation file in the MIT-BIH Arrhythmia database.
	\item Splitting the continuous ECG signal to a sequence of heartbeats based on the extracted $t$ waves and assigning a label to each heartbeat based on the annotation file. 
	\item Resizing each heartbeat to a predefined fixed length (280 samples).
\end{enumerate}
 We would like to note that these pre-processing steps for beat extraction are very simple and do not involve any form of filtering or noise removal methods.

\begin{figure*}[htb]
\centering
  \includegraphics[height=0.55\textheight,width=\linewidth]{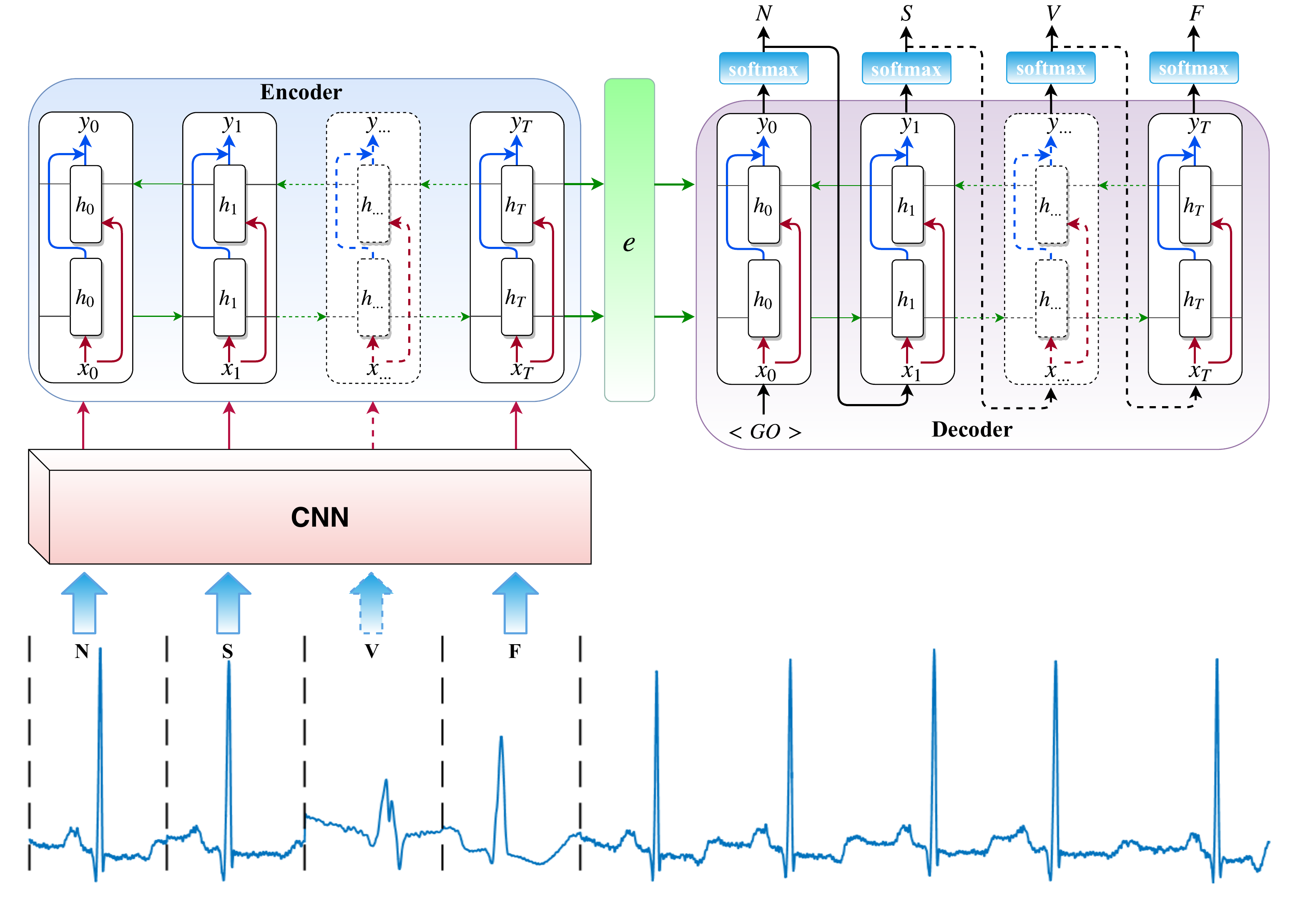}
  \caption{The proposed sequence to sequence deep learning network architecture for automatic heartbeat detection.} 
  \label{fig:final-model}
\end{figure*}

\subsection{The architecture}
The sequence to sequence models have shown very impassive results in neural machine translation applications, nearly similar to human-level performance \cite{johnson2016google}. 
The architecture of sequence to sequence networks is usually composed of two main parts of the recurrent neural network (RNN) encoder and decoder. In this study, we leverage an RNN sequence to sequence model along with a convolutional neural network (CNN) to perform a heartbeat detection task. \textbf{}

\begin{figure*}[htb]
\centering
  \includegraphics[height=0.3\textheight,width=\linewidth,keepaspectratio]{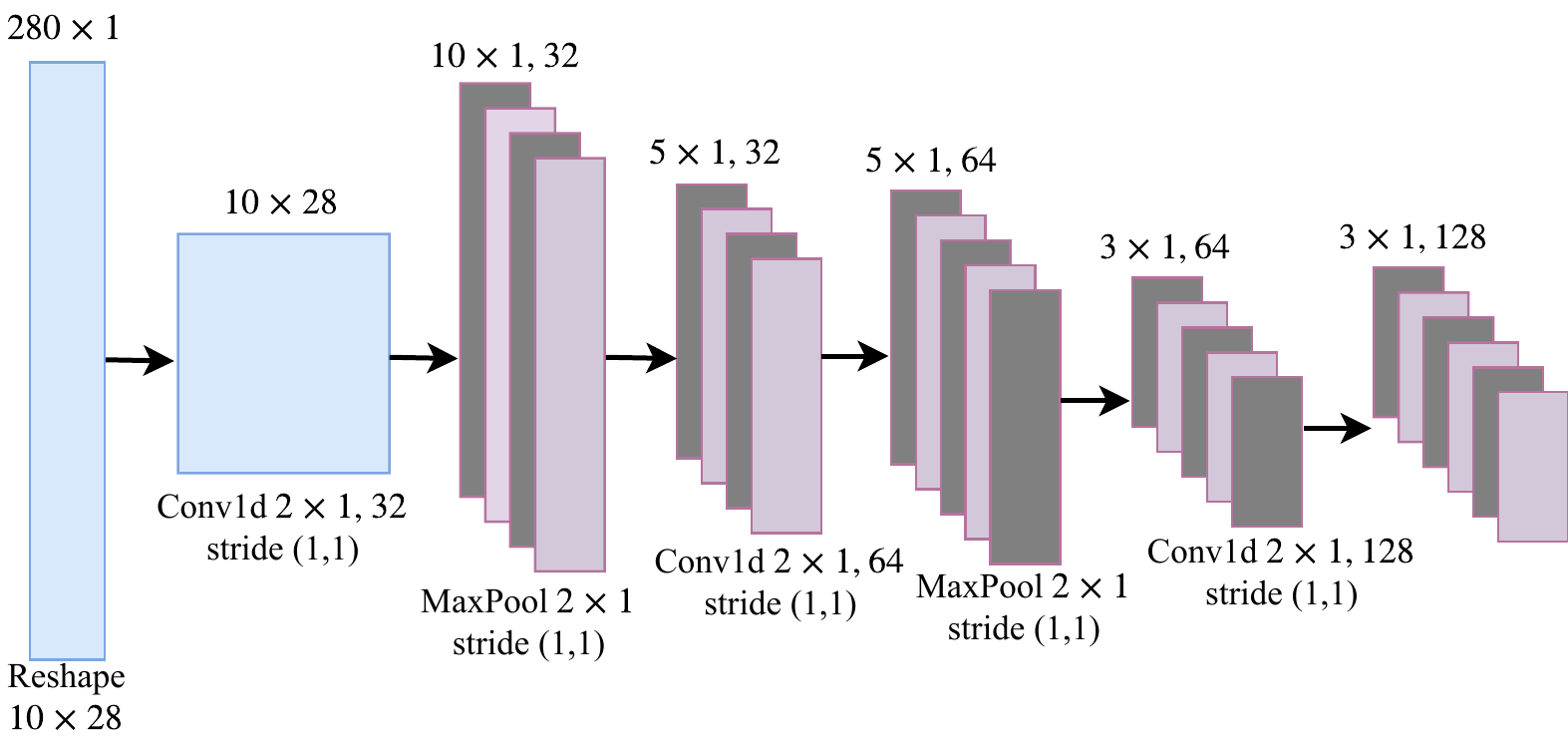}
  \caption{A diagram of convolutional layers used in the proposed model.} 
  \label{fig:cnn_part}
\end{figure*}

Fig. \ref{fig:final-model} illustrates the proposed network architecture  for automatic beat classification.
The CNN  consists of three consecutive one-dimensional convolutional layers. The first layer is composed of 32 1-D convolution filters with a kernel size of $2\times1$ and a stride 1, followed by a Rectified Linear Unit (ReLU) non-linearity. The second layer consists of 64 1-D convolution filters with a kernel size of $2\times1$ and a stride 1, again followed by an ReLU. Finally, the third layer is comprised of 128 1-D convolution filters with a kernel size of $2\times1$ and a stride 1, followed by a rectifier non-linearity. Each convolutional layer except the last layer is also followed by a max pooling layer of pooling region of size $2\times1$ with a stride 1. At each time-step of training/testing the model, a sequence (size of $maxtime$) of ECG beats is fed into the CNN in order for feature extraction. The last convolutional layer outputs
the $maxtime$ of $F$ feature maps of size $k\times1$ (e.g, here, we reached 128 feature maps $3\times1$). In the end, each beat of the input sequence is associated with a vector $c \in \mathbb{R}^d.$ Figure \ref{fig:cnn_part} depicts the detailed network. 

The sequence to sequence model is designed based on the encoder-decoder abstract ideas. The encoder encodes the input sequence, while the decoder computes the category of each beat of the input sequence. The encoder is actually composed of long short-term memory (LSTM) units, which is also called the many to one LSTM. The LSTMs can capture the complex and long short-term context dependencies between the inputs and the targets \cite{fernandez2015using}. This is due to the fact that they capture non-linear dependencies on entire observation when predicting a target. The (time) sequence of input feature vectors herein are fed to the LSTMs and then the last hidden state calculated by the LSTM is considered as the encoder representation and is used to initialize the fist hidden state of the decoder, as depicted in Fig. \ref{fig:final-model}.

We have utilized the bidirectional recurrent neural network (BiRNN) units in the network architecture instead of the standard LSTM (i.e., RNN). Standard RNNs are unidirectional, hence they are restricted to the use of the previous input state. To address this limitation, the BiRNN have been proposed \cite{schuster1997bidirectional}, which can process data in both forward and backward directions. Thus, the current state has access to previous and future input information simultaneously. The BiRNN consists of a forward network and a backward network. The input sequence is fed in normal time order, $t = 1, ...,T$ for the forward network, and in reverse time order, $t = T, ..., 1$ for the backward network. Finally, the weighted sum of the outputs of the two networks is computed as the output of the BiRNN. This mechanism can be formulated as follow:

\begin{align}
  &\begin{aligned} 
    \overrightarrow{h_t}= \tanh(\overrightarrow{W}x_t+\overrightarrow{V}{\overrightarrow{h}}_{t-1}+\overrightarrow{b})
  \end{aligned}\\
  &\begin{aligned}
   \overleftarrow{h_t}= \tanh(\overleftarrow{W}x_t+\overleftarrow{V}{\overleftarrow{h}}_{t+1}+\overleftarrow{b})
  \end{aligned} \\
    &\begin{aligned}
  y_t = (U[ \overrightarrow{h_t}; \overleftarrow{h_t}] + b_y),
  \end{aligned}
\end{align}
where ($\overrightarrow{h_t}$, $\overrightarrow{b}$) are the hidden state and the bias of the forward network, and ($\overleftarrow{h_t}$, $\overleftarrow{b}$) are the hidden state and the bias of the backward network. Also, $x_t$ and $y_t$ are the input and the output of the BiRNN, respectively.
The decoder is used to generate the target sequence beat by beat. Similar to the encoder, the building block of the decoder is an LSTM, but a many-to-many LSTM. The decoder gets a new representation of the input sequence generated by the encoder to initialize its hidden state. It also takes the same given target shifted by one and started with a special feature vector $<GO>$ as input. We should note that the input (the shifted target) is just used during the training phase and is not applied for the testing phase. Then, a softmax is applied to the output of the LSTM to convert it to a vector of probabilities $p \in \mathbb{R}^C$, where $C$ represents the number of categories (i.e., the heartbeat types) and each element of $p$ indicates the probability of each class in the category.

\section{Experimental evaluation}
\label{sec:experimental}
\subsection{Experimental setup}
\label{sec:setup}
The performance of the proposed heartbeat detection method has been evaluated using the MIT-BIH arrhythmia database using both inter-patient and intra-patient evaluation methods. Heartbeat category distribution of extracted beats from the MIT-BIH arrhythmia is not uniform and the numbers of normal beats are much greater than other categories. Machine learning approaches usually have trouble learning when one class dominates the others. To cope with this, the dataset has been over-sampled to nearly reach a balanced number of beats in each category (group). To this end, we used the synthetic minority over-sampling technique (SMOTE) which generates the synthetic data points by considering the similarities between existing minority samples \cite{chawla2002smote}.

Similar to several works in the literature, we evaluated the arrhythmia classifier on four cardiac cycles including N (normal and bundle branch block beats), S (supraventricular ectopic beats), V (ventricular ectopic beats),  and F (fusion of N and V beats) (for more details refer to Table \ref{tab:cat_beats}).
For the intra-patient scheme, we extracted totally 101,290 heartbeats, including 90,494 beats for the category N, 2,777 for the category S, 7,217 for the category V and 802 for the category F. 
To reduce the impact of class imbalance problem, the SMOTE technique was applied to upsample the categories with the small factions of heartbeats. We have trained the proposed model with 80\% of the dataset and evaluated it with the remaining 20\%. 
While for the inter-patient scheme, we had 45,796 beats of the category N, 941 beats of the category S, and 3,780 beats of the category V in DS1 (i.e., training set), and 44,196 beats for the category N, 1,836 beats for the category S, and 3,216 beats for the category V in DS2 (i.e., test set). 
Again,  the SMOTE approach was utilized to compensate the number of heartbeats of the categories with the lower heartbeats. 

The augmented data (i.e., synthetic data generated based on the original available data) should not be considered in the validation phase as the information of the validation set may exist in the training set. For instance, although the work reported in \cite{acharya2017deep}, followed this approach, we believe that the results are not quite accurate. 
It is worth mentioning that we generated new data samples by the SMOTE algorithm after splitting the heartbeats into the training and test datasets. Indeed, the oversampling was only performed on the training dataset for the intra-patient paradigm (and on DS1 for the inter-patient paradigm) and the information in the test dataset was not used to create synthetic data points. Therefore, the evaluation process is reliable and generalizable.

Four main measures usually were considered in the literature to evaluate the performance of heartbeat classification techniques including the sensitivity (SEN), positive predictive value (PPV), specificity (SPEC) (1 - False Positive Rate (FPR)) and accuracy (Acc) as defined in following:
\begin{flalign}
&SEN =TP/(TP+FN)& \\
&PPV = TP/(TP+FP)&\\
&SPEC = TN/(TN+FP)&\\
&Acc = (TP + TN)/(TN+FP+FP+FN),&
\end{flalign} 
where TP (True Positive), TN (True Negative), FP (False Positive) and FN (False Negative) indicate the number of heartbeats correctly labeled, number of heartbeats correctly identified as not correspond to the heartbeats, number of heartbeats that incorrectly labeled, and number of heartbeats which were not identified as the heartbeats that they should have been, respectively.

The network was trained for a maximum of 300 epochs and
the initial LSTM hidden and cell states were set to zero. All network weights were updated by the RMSProp algorithm  with mini batches of size 20 and a learning rate of $\alpha= 0.001$. 
\begin{table*} [ht]  
\caption{Intra-patient paradigm: Comparison of performance of the proposed heartbeat classifier against the state-of-the-art algorithms, considering randomly chosen sets for the training and testing based on the MIT-BIH arrhythmia database.}
 \centering{
\label{tab:compare-intra}
	\resizebox{1\linewidth}{!}{  
\begin{tabular}{ccccccccccccccccc}
\toprule
\textbf{Method} & \textbf{ACC} & \multicolumn{3}{c} {\textbf{N}} &  \multicolumn{3}{c} {\textbf{S}} &  \multicolumn{3}{c} {\textbf{V}} &  \multicolumn{3}{c} {\textbf{F}} &  \multicolumn{3}{c} {\textbf{Q}} \\
\cmidrule(lr){3-5} \cmidrule(lr){6-8}  \cmidrule(lr){9-11} \cmidrule(lr){12-14} \cmidrule(lr){15-17}
\textbf{} & \textbf{\%}  & \textbf{SEN} & \textbf{PPV}& \textbf{SPEC} &  \textbf{SEN} & \textbf{PPV}& \textbf{SPEC} &  \textbf{SEN} & \textbf{PPV}& \textbf{SPEC} &  \textbf{SEN} & \textbf{PPV}& \textbf{SPEC} & \textbf{SEN} & \textbf{PPV}& \textbf{SPEC} \\
\midrule
\textbf{Proposed method} & \textbf{99.92} 
& \textbf{1.00} & \textbf{99.86} &\textbf{98.87}
&\textbf{96.48}& \textbf{1.00}&\textbf{1.00}
&\textbf{99.50}&\textbf{99.79}&\textbf{99.98}
&\textbf{98.68}&\textbf{97.40}&\textbf{99.98} & -& -&- \\ 

\textbf{Kachuee et al. (2018)\cite{kachuee2018ecg}} & 93.4 & - &  - &-& -& -&-&-&-&-&-&-&- & -& -&- \\ 

\textbf{Acharya et al. (2017) \cite{acharya2017deep}} & 97.37 & 91.64 &  85.17 &96.01& 89.04& 94.76&98.77&94.07&95.08&98.74&95.21&94.69&98.67 & 97.39& 98.40&99.61 \\

\textbf{Ye et al. (2010) \cite{ye2010arrhythmia}} & 96.50 & 98.7 &  96.3 &-& 72.4& 94.5&-&82.6&97.8&-&65.6& 88.6&- & 95.8& 99.3&- \\ 
\textbf{Yu and Chou (2008) \cite{yu2008integration}} & 95.4 & 96.9 &97.3 & - & 73.8& 88.4&-
&92.3&94.3&-
&51.0&73.4&-
&94.1&80.8&- \\ 

\textbf{Song et al. (2005) \cite{song2005support}} & 98.7 & 99.5 & 98.9 &-& 86.4& 94.3&-&95.8&97.4&-&73.6&90.2&- & -& -&- \\ 
 
 \bottomrule  
\end{tabular} }
}
\end{table*}

\begin{table*}[ht] 
\caption{Inter-patient paradigm: Comparison of performance of the proposed heartbeat classifier against the state-of-the-art algorithms, considering DS1 as training dataset and DS2 as test dataset based on the MIT-BIH arrhythmia database.}
 \centering{
\label{tab:compare-inter}
	\resizebox{1\linewidth}{!}{  
\begin{tabular}{ccccccccccccccccc}
\toprule
\textbf{Method} & \textbf{ACC} & \multicolumn{3}{c} {\textbf{N}} &  \multicolumn{3}{c} {\textbf{S}} &  \multicolumn{3}{c} {\textbf{V}} &  \multicolumn{3}{c} {\textbf{F}} &  \multicolumn{3}{c} {\textbf{Q}} \\
\cmidrule(lr){3-5} \cmidrule(lr){6-8}  \cmidrule(lr){9-11} \cmidrule(lr){12-14} \cmidrule(lr){15-17}
\textbf{} & \textbf{\%}  & \textbf{SEN} & \textbf{PPV}& \textbf{SPEC} &  \textbf{SEN} & \textbf{PPV}& \textbf{SPEC} &  \textbf{SEN} & \textbf{PPV}& \textbf{SPEC} &  \textbf{SEN} & \textbf{PPV}& \textbf{SPEC} & \textbf{SEN} & \textbf{PPV}& \textbf{SPEC} \\
\midrule
\textbf{Proposed method} & \textbf{99.53} & \textbf{99.68} &  \textbf{99.55} &\textbf{96.05}
&\textbf{88.94}& \textbf{92.57}&\textbf{99.72}
&\textbf{99.94}&\textbf{99.50}&\textbf{99.97} &-&-&- & -& -&-
 \\ 

\textbf{Garcia et al. (2017)\cite{garcia2017inter}} 
& 92.4 & 94.0 &  98.0 &82.6& 62.0& 53.0&97.9
&87.3&59.4&95.9
&-&-&-&-&-&- \\
\textbf{Lin and Yang (2014) \cite{lin2014heartbeat}} & 93.0 & 91.0 & 99.0 &-
& 81.0&31.0&-
&86.0&73.0&-&-&-&- & -& -&- \\ 
\textbf{Ye et al. (2010) \cite{ye2010arrhythmia}} & 75.2 & 80.2 &  78.2 &-& 3.2& 10.3&-&50.2&48.5&-&-& -&- & -& -&- \\ 
\textbf{Yu and Chou (2008) \cite{yu2008integration}} & 75.2 & 78.3 &79.2 & - & 1.8& 5.9&-
&83.9&66.4&-
&0.3&0.1&-
&-&-&- \\ 

\textbf{Song et al. (2005) \cite{song2005support}} & 76.3 & 78.0 & 83.9 &-& 27.0& 48.3&-&80.8&38.7&-&-&-&- & -& -&- \\ 
 
 \bottomrule  
\end{tabular} }
}
\end{table*}
\subsection{Results and Discussion}
\label{sec:results}
The results are presented for two evaluation scenarios of intar-patient paradigm, in which the training and test sets were randomly chosen from all available patients' heartbeats, and the inter-patient paradigm, in which the training and test have been performed on the heartbeats of extracted from DS1 and DS2, respectively (i.e., no common individual in test and training sets). Table \ref{tab:compare-intra} presents a comparison of heartbeat classification results for the proposed method and the existing algorithms, considering intra-patient scheme. As it is clear, our sequence to sequence model outperforms all state-of-the-art algorithms significantly in terms of all evaluation metrics including overall accuracy, sensitivity, positive predictive value and specificity for the considered groups, N, S, V, F.
In addition, it is worth noticing that the work done by Acharya et al. \cite{acharya2017deep} used artificially augmented dataset to build the model, and their model evaluation was performed using augmented data, while we evaluated our model on real data samples without including any augmentations in the test set.

As confirmed by the results, our proposed method can provide a robust solution for class imbalance problem as one of the key challenges in dealing with medical data, which is due to the limited availability of abnormal classes compared to the normal classes in biomedical datasets. It is shown in Table \ref{tab:compare-intra} that our model achieves remarkable outcomes for the category F with only 802 heartbeats and the category S with 2,777 heartbeats.

We also validated our method using the more realistic evaluation method of inter-patient for heartbeat classifiers based on using DS1 set for training and DS2 set for testing  \cite{de2004automatic,luz2016ecg}. Table \ref{tab:compare-inter} presents the performance comparison between our proposed method and several state-of-the-art methods using MIT-BIH arrhythmia database where the inter-patient paradigm is considered. As it can be seen from the table, overall the proposed method has better performance for classifying all heartbeat categories compared to works listed in Table \ref{tab:compare-inter}. In spite of the low number of $S$ heartbeats in the training set, our proposed method obtained significant evaluation results.

The proposed method is generic in nature and it is expected to achieve a promising performance in several biomedical applications dealing with class imbalance problem. In the proposed classification structure, first, the CNN extracts a set of meaningful features of the given ECG heartbeats (Note, the model was trained by added synthetic data generated by SMOTE algorithm to the available samples to compensate the number of small categories like F). Then, the encoder maps the features to new feature representations, capturing temporal patterns, and finally, the decoder takes the feature representations and produces the outputs (i.e., the labels for each heartbeat of the input sequence), considering complex context dependencies between the inputs and the targets.

\section{Conclusion}
\label{sec:conc}
In this study, we presented a novel and effective automatic heartbeat classification/annotation, considering intar- and inter- patient schemes and validated its performance using the MIT-BIH arrhythmia database. The proposed method leverages the ability of deep convolutional neural network and encoder-decoder network in which we have used a bidirectional recurrent neural network as its building blocks. According to the results, the suggested method significantly outperforms the existing algorithms in the literature for both intar-patient paradigm and inter-patient paradigm. Furthermore, the proposed method can be applied to several biomedical applications such as sleep staging where there are strong dependencies between each stage and sufficient data are not available. Also, the proposed network with a low number of parameters (i.e., with a maximum size of 5.5MB) can be used with wearable devices.





\bibliographystyle{IEEEbib}
\bibliography{arxiv-inter}

\end{document}